# Meson Spectra using Nikiforov-Uvarov Method

**Hesham Mansour**[*] , **Ahmed Gamal**[1]

Physics Department, faculty of science, Cairo University, Giza, Egypt
Correspondence, Ahmed Gamal, Department of Physics, Cairo University, Giza, Egypt.

**A B S T R A C T**

There are many phenomenological potentials using different techniques to describe the spectroscopy of the quarkonium systems like ($c\bar{c}, b\bar{b}$ and $\bar{b}c$). In the present work we choose a phenomenological potential (linear, Yukawa and harmonic potentials) in the framework of the nonrelativistic Schrödinger's equation with relativistic corrections using Nikiforov-Uvarov (NU) method. We obtained the eigen function and eigen value comparing the results with the available experimental data. The results are in good agreement with the available experimental data .

## Introduction

Theorists have been trying to explain the behavior and characteristics of the quarkonium system by using different techniques and models like Bethe-Salpeter equation [1-7], lattice quantum chromodynamics techniques [8-11], The relativistic approaches [6,7,12-16], the semi-relativistic approaches [17-19], the nonrelativistic approaches [7,20-28] in which they used different phenomenological potentials such as the Coulomb and linear potentials [29-33], the Woods-Saxon Potential [26,34], power potentials [21,35,36], exponential potentials [32,33], harmonic and anharmonic potentials [12,22,28,37-39]. In the present work, we will use nonrelativistic Schrödinger equation and solve it by the Nikiforov-Uvarov (NU) method [10,22,28-33,37,38,40-50], with pheneomenlogical potentials (Yukawa, linear, harmonic). Relativistic corrections [14,20-22,27,31] are taken into account for the quarkonium systems. A fitting with the available experimental data was made to obtain the parameters of the potential.

---

[1] E-mail: alnabolci2010@gmail.com
[*] FInstP

## Theory

The masses of the quark and antiquark in the quarkonium system are bigger than chromodynamics scaling i.e. $M_{q,\bar{q}} \gg \Lambda_{QCD}$. So, non-relativistic treatment is suitable for heavy bound systems. Schrödinger equation of two-body system in spherical symmetric potential reads

$$\frac{d^2R}{dr^2} + \left[\frac{2\mu}{\hbar^2}(E - V(r)) - \frac{l(l+1)}{r^2}\right]R = 0 \qquad (1)$$

Where $\mu$ is reduced mass, $E$ is energy eigenvalue, $l$ is orbital quantum number and R (r) is a radial wavefunction solution of the Schrödinger equation. Our radial potential is taken as:

$$V(r) = \frac{-b}{r}e^{-cr} + ar + dr^2 \qquad (2)$$

Also, we use relativistic corrections (spin dependent splitting) terms: spin-spin interaction $V_{S-S}(r)$, spin-orbital interaction $V_{S-l}(r)$, and tensor interaction $V_T(r)$.

$$V_{S-l-T}(r) = V_{S-S}(r) + V_{S-l}(r) + V_T(r) \qquad (3)$$

Where

$$V_{S-S}(r) = \frac{2}{3m_q m_{\bar{q}}} \nabla^2 V_V(r)[\vec{S_q} \cdot \vec{S_{\bar{q}}}] \qquad (4)$$

$$V_{S-l}(r) = \frac{1}{2m_q m_{\bar{q}} r}\left[3\frac{dV_V(r)}{dr} - \frac{dV_s(r)}{dr}\right][\vec{L} \cdot \vec{S}] \qquad (5)$$

$$V_T(r) = \frac{1}{12 m_q m_{\bar{q}}}\left[\frac{1}{r}\frac{dV_V(r)}{dr} - \frac{d^2V_V(r)}{dr^2}\right]\left[6\left(\vec{S_q}\cdot\frac{\vec{r}}{|r|}\right)\left(\vec{S_{\bar{q}}}\cdot\frac{\vec{r}}{|r|}\right) - 2\vec{S_q}\cdot\vec{S_{\bar{q}}}\right] \qquad (6)$$

$V_V$ is a vector potential term and $V_s$ is a scalar potential one.

The total potential $V_{tot}(r)$ due to spin splitting terms is

$$V_{tot}(r) = V(r) + V_{s-l-T}(r) \qquad (7)$$

using equations (4), (5), (6) in equation (2), we get

$$V_{tot}(r) = \frac{e^{-cr}}{r^3}\left[\frac{3bv(ls)}{2m_q m_{\bar{q}}} + \frac{bv(t)}{4m_q m_{\bar{q}}}\right] + \frac{e^{-cr}}{r^2}\left[\frac{3cbv(ls)}{2m_q m_{\bar{q}}} + \frac{cbv(t)}{4m_q m_{\bar{q}}}\right]$$
$$+ \frac{e^{-cr}}{r}\left[\frac{-2c^2 bv(ss)}{3m_q m_{\bar{q}}} + \frac{c^2 bv(t)}{12 m_q m_{\bar{q}}} - b\right]$$
$$+ \frac{1}{r}\left[\frac{4a_v v(ss)}{3m_q m_{\bar{q}}} + \frac{(3a_v - a_S)v(ls)}{2m_q m_{\bar{q}}} + \frac{a_v v(t)}{12 m_q m_{\bar{q}}}\right] - \frac{dv(ls)}{m_q m_{\bar{q}}} + ar$$
$$+ dr^2 \qquad (8)$$

Where

$$v(ss) = [\vec{S_q} \cdot \vec{S_{\bar{q}}}], v(ls) = [\vec{L} \cdot \vec{S}],$$
$$v(T) = -2\left[\vec{S_q} \cdot \vec{S_{\bar{q}}} - 3\left(\vec{S_q} \cdot \frac{\vec{r}}{|r|}\right)\left(\vec{S_{\bar{q}}} \cdot \frac{\vec{r}}{|r|}\right)\right], \quad a_S + a_v = a \qquad (9)$$

By using natural units and substitute equation (8) into equation (1), we obtain

$$\frac{d^2 R}{dr^2} + \left[2\mu E - \frac{2\mu f_1 e^{-cr}}{r^3} - \frac{2\mu f_2 e^{-cr}}{r^2} - \frac{2\mu f_3 e^{-cr}}{r} - \frac{2\mu f_4}{r} + 2\mu f_5 - 2\mu ar - 2\mu dr^2 \right.$$
$$\left. - \frac{l(l+1)}{r^2}\right] R = 0 \qquad (10)$$

Where

$$f_1 = \frac{3bv(ls)}{2m_q m_{\bar{q}}} + \frac{bv(t)}{4m_q m_{\bar{q}}} = \frac{b}{4m_q m_{\bar{q}}}[6v(ls) + v(t)], \qquad f_5 = \frac{dv(ls)}{m_q m_{\bar{q}}} \qquad (11)$$

$$f_2 = \frac{3cbv(ls)}{2m_q m_{\bar{q}}} + \frac{cbv(t)}{4m_q m_{\bar{q}}} = \frac{bc}{4m_q m_{\bar{q}}}[6v(ls) + v(t)] = cf_1 \qquad (12)$$

$$f_3 = \frac{-2c^2 bv(ss)}{3m_q m_{\bar{q}}} + \frac{c^2 bv(t)}{12 m_q m_{\bar{q}}} - b = \frac{bc^2}{12 m_q m_{\bar{q}}}[-8v(ss) + v(t)] - b \qquad (13)$$

$$f_4 = \frac{4a_v v(ss)}{3m_q m_{\bar{q}}} + \frac{(3a_v - a_S)v(ls)}{2m_q m_{\bar{q}}} + \frac{a_v v(t)}{12 m_q m_{\bar{q}}}$$
$$= \frac{1}{12 m_q m_{\bar{q}}}[16 a_v v(ss) + 6(3a_v - a_S)v(ls) + a_v v(t)] \qquad (14)$$

We transform the exponential form into summation by Taylor series around zero

$$e^{-cr} = \sum_{i=0}^{i} \frac{(-1)^i c^i}{i!} r^i \qquad (15)$$

substituting equation (15) in equation (10) and rearrange it, we get

$$\frac{d^2R}{dr^2} + \left[ h_4 - \frac{h_1}{r^3} + \frac{h_2}{r^2} + \frac{h_3}{r} - Ar - Dr^2 + \sum_{i=0}^{i} M_i \, r^i \right] R = 0 \qquad (16)$$

Where

$$\epsilon = 2\mu E, \quad 2\mu f_1 = h_1 \,, \quad 2\mu f_1 c - 2\mu f_2 - l(l+1) = h_2 \qquad (17)$$

$$2\mu f_2 c - \frac{2\mu f_1 c^2}{2} - 2\mu f_3 - 2\mu f_4 = h_3, \quad 2\mu f_5 + \epsilon = h_4, \quad 2\mu d = D \qquad (18)$$

$$2\mu a = A, \quad 2\mu \frac{(-1)^i c^{i+1}}{(i+1)!} \left[ \frac{-f_1 c^2}{(i+3)} + f_3 \right] = M_i \qquad (19)$$

let $x = \frac{1}{r}$ and by substituting in equation (16), we get

$$\frac{d^2R}{dx^2} + \frac{2}{x}\frac{dR}{dx} + \frac{1}{x^2}\left[ \frac{h_4}{x^2} - \frac{D}{x^4} - \frac{A}{x^3} - h_1 x + h_2 + \frac{h_3}{x} + \sum_{i=0}^{i} M_i \, x^{-i-2} \right] R = 0 \qquad (20)$$

Due to the singularity point in equation (20), we put $y + \delta = x$, and by using the Taylor's series to expand to second order terms, one obtains

$$\frac{d^2R}{dx^2} + \frac{2}{x}\frac{dR}{dx} + \frac{1}{x^2}[-q + wx - zx^2]R = 0 \qquad (21)$$

Where

$$-q = \sum_{i=0}^{i} \frac{M_i \, (i+4)(i+3)}{2\delta^{i+2}} + h_2 + \frac{3h_3}{\delta} + \frac{6h_4}{\delta^2} - \frac{15D}{\delta^4} - \frac{10A}{\delta^3} \qquad (22)$$

$$-w = \sum_{i=0}^{i} \frac{M_i \, (i+2)(i+4)}{\delta^{i+3}} + h_1 + \frac{3h_3}{\delta^2} + \frac{8h_4}{\delta^3} - \frac{24D}{\delta^5} - \frac{15A}{\delta^4} \qquad (23)$$

$$-z = \sum_{i=0}^{i} \frac{M_i \, (i+2)(i+3)}{2\delta^{i+4}} + \frac{h_3}{\delta^3} + \frac{3h_4}{\delta^4} - \frac{10D}{\delta^6} - \frac{6A}{\delta^5} \qquad (24)$$

One can use the Nikiforov-Uvarov method (NU) for equation (21) to get eigenvalue and eigenfunction equations. The eigenfunction equation is

$$R(r) = N_{nls} r^{1+\sqrt{\left(q+\frac{1}{4}\right)}} e^{\frac{\sqrt{z}}{r}} L_n^{1+2\sqrt{\left(q+\frac{1}{4}\right)}} \left( \frac{2\sqrt{z}}{r} \right) \qquad (25)$$

where $L_n^{2\sqrt{\left(q+\frac{1}{4}\right)}}\left(\frac{2\sqrt{z}}{r}\right)$ is the Rodrigues's formula of the associated Laguerre polynomial and $N_{nls}$ is a normalization constant.

The eigen value equation is

$$E = E_1 - \frac{1}{6\mu}\left[\frac{E_2}{2\acute{n} + 1 + 2\sqrt{E_3 + \left(\acute{l}+\frac{1}{2}\right)^2}}\right]^2 \tag{26}$$

Where

$$E_1 = \frac{10d}{3\delta^2} + \frac{2a}{\delta} - f_5 - \frac{\left(f_2 c - \frac{f_1 c^2}{2} - f_3 - f_4\right)\delta}{3}$$
$$-\frac{1}{6}\sum_{i=0}^{i}\frac{(-1)^i c^{i+1}(i+2)(i+3)}{(i+1)!\,\delta^i}\left[\frac{-f_1 c^2}{(i+3)} + f_3\right] \tag{27}$$

$$E_2 = \frac{24D}{\delta^3} + \frac{15A}{\delta^2}$$
$$-2\mu\sum_{i=0}^{i}\frac{(-1)^i c^{i+1}(i+2)(i+4)}{(i+1)!\,\delta^{i+1}}\left[\frac{-f_1 c^2}{(i+3)} + f_3\right] - 3h_3$$
$$-\frac{8(2\mu f_5 + \epsilon)}{\delta} - h_1\delta^2 \tag{28}$$

$$E_3 = \frac{15D}{\delta^4} + \frac{10A}{\delta^3} - 2\mu\sum_{i=0}^{i}\frac{(-1)^i c^{i+1}(i+4)(i+3)}{2(i+1)!\,\delta^{i+2}}\left[\frac{-f_1 c^2}{(i+3)} + f_3\right] - \frac{3h_3}{\delta}$$
$$-\frac{6(2\mu f_5 + \epsilon)}{\delta^2} \tag{29}$$

$\acute{n}$ is not a principal quantum number, but it is a dimensional parameter and has $n$ principle quantum number. Explicitly
$$\acute{n} = (n + 0.5)e - 0.5 \quad and \quad e = 1\ GeV$$

Similarly, $\acute{l}$ is not an orbital quantum number but it is a dimensional parameter and has $l$ orbital quantum number. Explicitly
$$\acute{l} = (l + 0.5)e - 0.5 \quad and \quad e = 1\ GeV$$

We use equation (15) to change the form of equations (29), (28), (27). So, the mass spectra become

$$M(q\bar{q}) = E_1 + m_q + m_{\bar{q}} - \frac{1}{6\mu}\left[\frac{E_2}{2n + 1 + 2\sqrt{E_3 + \left(l + \frac{1}{2}\right)^2}}\right]^2 \tag{30}$$

Where

$$E_1 = \frac{10d}{3}r_0^2 + 2(a_v + a_s)r_0 - \frac{dv(ls)}{m_q m_{\bar{q}}}$$

$$+ \frac{1}{36 m_q m_{\bar{q}} r_0}[16 a_v v(ss) + 6(3a_v - a_s)v(ls) + a_v v(t)]$$

$$- \frac{bce^{-cr_0}}{6}\left[\frac{[v(t) - 8v(ss)]c^2}{12 m_q m_{\bar{q}}} - 1\right]\left[4 - cr_0 - \frac{2}{cr_0}\right]$$

$$+ \frac{bc^2 e^{-cr_0}}{24 m_q m_{\bar{q}}}[6v(ls) + v(t)]\left[c - \frac{1}{r_0}\right] \tag{31}$$

$$E_2 = 2\mu\Bigg[24 d r_0^3 + 15(a_v + a_s)r_0^2 + \frac{[16 a_v v(ss) + 6(3a_v - a_s)v(ls) + a_v v(t)]}{4 m_q m_{\bar{q}}}$$

$$- 8\left[E + \frac{dv(ls)}{m_q m_{\bar{q}}}\right] r_0$$

$$- e^{-cr_0} b\left[\frac{[-8v(ss) + v(t)]c^2}{12 m_q m_{\bar{q}}} - 1\right][5cr_0 - c^2 r_0^2 - 3]$$

$$+ \frac{bc^2 e^{-cr_0}[6v(ls) + v(t)]}{4 m_q m_{\bar{q}}}\left[cr_0 - 2 - \frac{1}{r_0 c} - \frac{1}{c^2 r_0^2}\right]\Bigg] \tag{32}$$

$$E_3 = 2\mu\Bigg[15 d r_0^4 + 10(a_v + a_s)r_0^3$$

$$+ \frac{r_0}{4 m_q m_{\bar{q}}}[16 a_v v(ss) + 6(3a_v - a_s)v(ls) + a_v v(t)]$$

$$- 6 r_0^2\left[E + \frac{dv(ls)}{m_q m_{\bar{q}}}\right]$$

$$- \frac{r_0 e^{-cr_0} b}{2}\left[\frac{[-8v(ss) + v(t)]c^2}{12 m_q m_{\bar{q}}} - 1\right][6cr_0 - c^2 r_0^2 - 6]$$

$$+ \frac{bc^2 r_0 e^{-cr_0}}{8 m_q m_{\bar{q}}}[6v(ls) + v(t)][cr_0 - 3]\Bigg] \tag{33}$$

The energy eigenvalue equation (30) has spin-orbital-tensor coefficients $v(ss), v(sl), v(T)$ [51-67]. Also, it has potential parameters ($a_s, a_v, b, d, c$) and $r_0$ due to the expansion, so we have six parameters of the eigenvalue equation which can be obtained from the experimental data by best fitting.

## Results and Discussions

In table (1), potential parameters are shown for the systems under consideration. We use spectroscopic notation for the levels ($n^{2S+1}L_J$). S is total spin of the system, L is the orbital quantum number, n is the principal quantum number, J is the total (orbital + spin) quantum number. By utilizing equation (30) and table (1), we get the mass spectra of the different quantum states (present work or the theoretical states) as shown in the tables (2-7). We notice from the tables that the average of the chi-squared values is 0.000793 for the charmonium system, 0.00012 for the bottomonium system and 0.00013 for the Bc meson system. Besides of the small percentage errors (between two brackets) in the last column the experimental values are given(PDG [68]). Our work is better in comparison with available experimental than other values in the literatures [1,10,16,17,27,59,61,63,70-76].

Table 1. Parameter values for each system

| Variables / Systems | $m_q$ | $m_{\bar{q}}$ | $r_0$ | as | av | b | c | d |
|---|---|---|---|---|---|---|---|---|
| Units | GeV | GeV | GeV$^{-1}$ | GeV$^2$ | GeV$^2$ | -- | GeV | GeV$^3$ |
| c quark | 1.317 | 1.317 | 28.6261 | 0.2572 | 0.07565 | 433.7577 | 0.07217 | -0.0056 |
| b quark | 4.584 | 4.584 | | | | | | |

Table 2. Charmonia mass spectrum of S and P-states in GeV

| Level | Present work | [1] | [69] | [70] | [27] | [71] | [17] | [72] | [73] | [10] | [59] | PDG [68] |
|---|---|---|---|---|---|---|---|---|---|---|---|---|
| $1^1S_0$ | 3.0413 | 2.93 | 2.981 | 2.984 | 2.989 | 2.979 | 2.980 | 2.980 | 2.982 | 3.088 | 2.979 | 2.984 (1.92%) |
| $1^3S_1$ | 3.1404 | 3.11 | 3.096 | 3.097 | 3.094 | 3.097 | 3.097 | 3.097 | 3.090 | 3.168 | 3.096 | 3.097 (1.39%) |
| $2^1S_0$ | 3.6610 | 3.68 | 3.635 | 3.637 | 3.602 | 3.623 | 3.597 | 3.633 | 3.630 | 3.669 | 3.600 | 3.639 (0.61%) |
| $2^3S_1$ | 3.7017 | 3.68 | 3.685 | 3.679 | 3.681 | 3.673 | 3.685 | 3.690 | 3.672 | 3.707 | 3.680 | 3.686 (0.43%) |
| $3^1S_0$ | 4.1347 | -- | 3.989 | 4.004 | 4.058 | 3.991 | 4.014 | 3.992 | 4.043 | 4.067 | 4.011 | -- |

| State | | | | | | | | | | | |
|---|---|---|---|---|---|---|---|---|---|---|---|
| $3^3S_1$ | 4.0502 | 3.80 | 4.039 | 4.030 | 4.129 | 4.022 | 4.095 | 4.030 | 4.072 | 4.094 | 4.077 | 4.039 (0.28%) |
| $4^1S_0$ | 4.4136 | -- | 4.401 | 4.264 | 4.448 | 4.250 | 4.433 | 4.244 | 4.384 | 4.398 | 4.397 | -- |
| $4^3S_1$ | 4.4185 | -- | 4.427 | 4.281 | 4.514 | 4.273 | 4.477 | 4.273 | 4.406 | 4.420 | 4.454 | 4.421 (0.06%) |
| $5^1S_0$ | 4.6618 | -- | 4.811 | 4.459 | 4.799 | 4.446 | -- | 4.440 | -- | -- | -- | -- |
| $5^3S_1$ | 4.6591 | -- | 4.837 | 4.472 | 4.863 | 4.463 | -- | 4.464 | -- | -- | -- | -- |
| $6^1S_0$ | 4.8825 | -- | 5.155 | -- | 5.124 | 4.595 | -- | 4.601 | -- | -- | -- | -- |
| $6^3S_1$ | 4.8801 | -- | 5.167 | -- | 5.185 | 4.608 | -- | 4.621 | -- | -- | -- | -- |
| $1^3P_0$ | 3.4137 | 3.32 | 3.413 | 3.415 | 3.428 | 3.433 | 3.416 | 3.392 | 3.424 | 3.448 | 3.488 | 3.415 (0.04%) |
| $1^3P_1$ | 3.5036 | 3.49 | 3.511 | 3.521 | 3.468 | 3.510 | 3.508 | 3.491 | 3.505 | 3.520 | 3.514 | 3.511 (0.21%) |
| $1^1P_1$ | 3.5180 | 3.43 | 3.525 | 3.526 | 3.470 | 3.519 | 3.527 | 3.524 | 3.516 | 3.536 | 3.536 | 3.525 (0.20%) |
| $1^3P_2$ | 3.4888 | 3.55 | 3.555 | 3.553 | 3.480 | 3.556 | 3.558 | 3.570 | 3.556 | 3.564 | 3.565 | 3.556 (0.25%) |
| $2^3P_0$ | 3.7646 | 3.83 | 3.870 | 3.848 | 3.897 | 3.842 | 3.844 | 3.845 | 3.852 | 3.870 | 3.947 | 3.918 (0.51%) |
| $2^3P_1$ | 3.8072 | 3.67 | 3.906 | 3.914 | 3.938 | 3.901 | 3.940 | 3.902 | 3.925 | 3.934 | 3.972 | -- |
| $2^1P_1$ | 3.8239 | 3.75 | 3.926 | 3.916 | 3.943 | 3.908 | 3.960 | 3.922 | 3.934 | 3.950 | 3.996 | -- |
| $2^3P_2$ | 3.9151 | -- | 3.949 | 3.937 | 3.955 | 3.937 | 3.994 | 3.949 | 3.972 | 3.976 | 4.021 | 3.927 (0.30%) |
| $3^3P_0$ | 4.0804 | -- | 4.301 | 4.146 | 4.296 | 4.131 | -- | 4.192 | 4.202 | 4.214 | -- | -- |
| $3^3P_1$ | 4.1210 | 3.91 | 4.319 | 4.192 | 4.338 | 4.178 | -- | 4.178 | 4.271 | 4.275 | -- | -- |
| $3^1P_1$ | 4.1368 | -- | 4.337 | 4.193 | 4.344 | 4.184 | -- | 4.137 | 4.279 | 4.291 | -- | -- |
| $3^3P_2$ | 4.1514 | -- | 4.354 | 4.211 | 4.358 | 4.208 | -- | 4.212 | 4.317 | 4.316 | -- | -- |
| $4^3P_0$ | 4.3621 | -- | 4.698 | -- | 4.653 | -- | -- | -- | -- | -- | -- | -- |
| $4^3P_1$ | 4.4005 | -- | 4.728 | -- | 4.696 | -- | -- | -- | -- | -- | -- | -- |
| $4^1P_1$ | 4.4155 | -- | 4.744 | -- | 4.704 | -- | -- | -- | -- | -- | -- | -- |
| $4^3P_2$ | 4.4298 | -- | 4.763 | -- | 4.718 | -- | -- | -- | -- | -- | -- | -- |

| Level | | | | | | | | | | | |
|---|---|---|---|---|---|---|---|---|---|---|---|
| $5^3P_0$ | 4.6129 | -- | -- | -- | 4.983 | -- | -- | -- | -- | -- | -- |
| $5^3P_1$ | 4.6493 | -- | -- | -- | 5.026 | -- | -- | -- | -- | -- | -- |
| $5^1P_1$ | 4.6636 | -- | -- | -- | 5.034 | -- | -- | -- | -- | -- | -- |

Table 3. Mass spectrum of the charmonia for D and F-states in GeV.

| Level | Present work | [69] | [72] | [73] | [70] | [1] | [27] | [71] | [17] | [10] | [59] |
|---|---|---|---|---|---|---|---|---|---|---|---|
| $1^3D_3$ | 3.51402 | 3.813 | 3.844 | 3.806 | 3.808 | 3.869 | 3.755 | 3.799 | 3.831 | 3.809 | 3.798 |
| $1^1D_2$ | 3.47795 | 3.807 | 3.802 | 3.799 | 3.805 | 3.739 | 3.765 | 3.796 | 3.824 | 3.803 | 3.796 |
| $1^3D_2$ | 3.46047 | 3.795 | 3.788 | 3.800 | 3.807 | 3.550 | 3.772 | 3.798 | 3.824 | 3.804 | 3.794 |
| $1^3D_1$ | 3.40228 | 3.783 | 3.729 | 3.785 | 3.792 | -- | 3.775 | 3.787 | 3.804 | 3.789 | 3.792 |
| $2^3D_3$ | 3.863 | 4.220 | 4.132 | 4.167 | 4.112 | 3.806 | 4.176 | 4.103 | 4.202 | 4.167 | 4.425 |
| $2^1D_2$ | 3.82825 | 4.196 | 4.105 | 4.158 | 4.108 | -- | 4.182 | 4.099 | 4.191 | 4.158 | 4.224 |
| $2^3D_2$ | 3.81161 | 4.190 | 4.095 | 4.158 | 4.109 | -- | 4.188 | 4.100 | 4.189 | 4.159 | 4.223 |
| $2^3D_1$ | 3.75606 | 4.105 | 4.057 | 4.142 | 4.095 | -- | 4.188 | 4.089 | 4.164 | 4.143 | 4.222 |
| $3^3D_3$ | 4.17431 | 4.574 | 4.351 | -- | 4.340 | -- | 4.549 | 4.331 | -- | -- | -- |
| $3^1D_2$ | 4.14085 | 3.549 | 4.330 | -- | 4.336 | -- | 4.553 | 4.326 | -- | -- | -- |
| $3^3D_2$ | 4.12502 | 4.544 | 4.322 | -- | 4.337 | -- | 4.557 | 4.327 | -- | -- | -- |
| $3^3D_1$ | 4.07208 | 4.507 | 4.293 | -- | 4.324 | -- | 4.555 | 4.317 | -- | -- | -- |
| $4^3D_3$ | 4.45154 | 4.920 | 4.526 | -- | -- | -- | 4.890 | -- | -- | -- | -- |
| $4^1D_2$ | 4.41933 | 4.898 | 4.509 | -- | -- | -- | 4.892 | -- | -- | -- | -- |
| $4^3D_2$ | 4.40431 | 4.896 | 4.504 | -- | -- | -- | 4.896 | -- | -- | -- | -- |
| $4^3D_1$ | 4.35388 | 4.857 | 4.480 | -- | -- | -- | 4.891 | -- | -- | -- | -- |
| $1^3F_2$ | 3.38961 | 4.041 | -- | 4.029 | -- | -- | 3.990 | -- | 4.068 | -- | -- |
| $1^3F_3$ | 3.46746 | 4.068 | -- | 4.029 | -- | 3.999 | 4.012 | -- | 4.070 | -- | -- |
| $1^1F_3$ | 3.48494 | 4.071 | -- | 4.026 | -- | 4.037 | 4.017 | -- | 4.066 | -- | -- |
| $1^3F_4$ | 3.54185 | 4.093 | -- | 4.021 | -- | -- | 4.036 | -- | 4.062 | -- | -- |
| $2^3F_2$ | 3.74376 | 4.361 | -- | 4.351 | -- | -- | 4.378 | -- | -- | -- | -- |
| $2^3F_3$ | 3.81815 | 4.400 | -- | 3.352 | -- | -- | 4.396 | -- | -- | -- | -- |
| $2^1F_3$ | 3.83479 | 4.406 | -- | 4.350 | -- | -- | 4.400 | -- | -- | -- | -- |
| $2^3F_4$ | 3.88949 | 4.434 | -- | 4.348 | -- | -- | 4.415 | -- | -- | -- | -- |
| $3^3F_2$ | 4.06014 | -- | -- | -- | -- | -- | 4.730 | -- | -- | -- | -- |
| $3^3F_3$ | 4.13113 | -- | -- | -- | -- | -- | 4.746 | -- | -- | -- | -- |
| $3^1F_3$ | 4.14695 | -- | -- | -- | -- | -- | 4.749 | -- | -- | -- | -- |

| Level | | | | | | | | | | |
|---|---|---|---|---|---|---|---|---|---|---|
| $3^3F_4$ | 4.1995 | -- | -- | -- | -- | -- | 4.761 | -- | -- | -- | -- |

Table 4. Mass spectrum of the bottomonia for S and P-states in GeV.

| Level | Present work | [74] | [69] | [70] | [1] | [77] | [63] | [72] | [75] | PDG [68] |
|---|---|---|---|---|---|---|---|---|---|---|
| $1^1S_0$ | 9.43601 | 9.402 | 9.398 | 9.390 | 9.414 | 9.389 | 9.393 | 9.392 | 9.455 | 9.398 (0.41%) |
| $1^3S_1$ | 9.49081 | 9.465 | 9.460 | 9.460 | 9.490 | 9.460 | 9.460 | 9.460 | 9.502 | 9.460 (0.33%) |
| $2^1S_0$ | 9.99146 | 9.976 | 9.990 | 9.990 | 9.987 | 9.987 | 9.987 | 9.991 | 9.990 | 9.999 (0.075%) |
| $2^3S_1$ | 10.01257 | 10.003 | 10.023 | 10.015 | 10.089 | 10.016 | 10.023 | 10.024 | 10.015 | 10.023 (0.1%) |
| $3^1S_0$ | 10.1386 | 10.336 | 10.329 | 10.326 | -- | 10.330 | 10.345 | 10.323 | 10.330 | -- |
| $3^3S_1$ | 10.32775 | 10.354 | 10.355 | 10.343 | 10.327 | 10.351 | 10.364 | 10.346 | 10.349 | 10.355 (0.26%) |
| $4^1S_0$ | 10.3236 | 10.523 | 10.573 | 10.584 | -- | 10.595 | 10.623 | 10.558 | -- | -- |
| $4^3S_1$ | 10.5461 | 10.635 | 10.586 | 10.597 | -- | 10.611 | 10.643 | 10.575 | 10.607 | 10.579 (0.31%) |
| $5^1S_0$ | 10.4977 | 10.869 | 10.851 | 10.800 | -- | 10.817 | -- | 10.741 | -- | -- |
| $5^3S_1$ | 10.82628 | 10.878 | 10.869 | 10.811 | -- | 10.831 | -- | 10.755 | 10.818 | 10.876 (0.46%) |
| $6^1S_0$ | 10.6615 | 11.097 | 11.061 | 10.997 | -- | 11.011 | -- | 10.892 | -- | -- |
| $6^3S_1$ | 10.97061 | 11.102 | 11.088 | 10.988 | -- | 10.988 | -- | 10.904 | 10.995 | 11.019 (0.44%) |
| $1^3P_0$ | 9.8432 | 9.847 | 9.859 | 9.864 | 9.815 | 9.865 | 9.861 | 9.862 | 9.855 | 9.859 (0.16%) |
| $1^3P_1$ | 9.87371 | 9.876 | 9.892 | 9.903 | 9.842 | 9.897 | 9.891 | 9.888 | 9.874 | 9.893 (0.195%) |
| $1^1P_1$ | 9.87919 | 9.882 | 9.900 | 9.909 | 9.806 | 9.903 | 9.900 | 9.896 | 9.879 | 9.899 (0.2%) |
| $1^3P_2$ | 9.89083 | 9.897 | 9.912 | 9.921 | 9.906 | 9.918 | 9.912 | 9.908 | 9.886 | 9.912 (0.214%) |
| $2^3P_0$ | 10.19625 | 10.226 | 10.233 | 10.220 | 10.254 | 10.226 | 10.230 | 10.241 | 10.221 | 10.232 (0.35%) |
| $2^3P_1$ | 10.21695 | 10.246 | 10.255 | 10.249 | 10.120 | 10.251 | 10.255 | 10.256 | 10.236 | 10.255 (0.37%) |
| $2^1P_1$ | 10.22153 | 10.250 | 10.260 | 10.254 | 10.154 | 10.256 | 10.262 | 10.261 | 10.240 | 10.260 (0.38%) |

| Level | Present work | [74] | [69] | [70] | [1] | [77] | [63] | [72] | [75] | PDG [68] |
|---|---|---|---|---|---|---|---|---|---|---|
| $2^3P_2$ | 10.22961 | 10.261 | 10.268 | 10.264 | -- | 10.269 | 10.271 | 10.268 | 10.246 | 10.269 (0.38%) |
| $3^3P_0$ | 10.1342 | 10.552 | 10.521 | 10.490 | -- | 10.502 | -- | 10.511 | 10.500 | -- |
| $3^3P_1$ | 10.1378 | 10.538 | 10.541 | 10.515 | 10.303 | 10.524 | -- | 10.507 | 10.513 | -- |
| $3^1P_1$ | 4.14695 | 10.541 | 10.544 | 10.519 | -- | 10.529 | -- | 10.497 | 10.516 | -- |
| $3^3P_2$ | 10.1405 | 10.550 | 10.550 | 10.528 | -- | 10.540 | -- | 10.516 | 10.521 | -- |
| $4^3P_0$ | 10.3193 | 10.775 | 10.781 | -- | -- | 10.732 | -- | -- | -- | -- |
| $4^3P_1$ | 10.3229 | 10.788 | 10.802 | -- | -- | 10.753 | -- | -- | -- | -- |
| $4^1P_1$ | 10.3242 | 10.790 | 10.804 | -- | -- | 10.757 | -- | -- | -- | -- |
| $4^3P_2$ | 10.3255 | 10.798 | 10.812 | -- | -- | 10.767 | -- | -- | -- | -- |
| $5^3P_0$ | 10.4936 | 11.004 | -- | -- | -- | 10.933 | -- | -- | -- | -- |
| $5^3P_1$ | 10.497 | 11.014 | -- | -- | -- | 10.951 | -- | -- | -- | -- |
| $5^1P_1$ | 10.4983 | 11.016 | -- | -- | -- | 10.955 | -- | -- | -- | -- |

Table 5. Mass spectrum of the bottomonia for D and F-states in GeV.

| Level | Present work | [74] | [69] | [70] | [1] | [77] | [63] | [72] | [75] | PDG [68] |
|---|---|---|---|---|---|---|---|---|---|---|
| $1^3D_3$ | 9.73855 | 10.115 | 10.166 | 10.157 | 10.232 | 10.156 | 10.163 | 10.177 | 10.127 | -- |
| $1^1D_2$ | 9.7355 | 10.148 | 10.163 | 10.153 | 10.194 | 10.152 | 10.158 | 10.166 | 10.123 | -- |
| $1^3D_2$ | 10.1126 | 10.147 | 10.161 | 10.153 | 10.145 | 10.151 | 10.157 | 10.162 | 10.122 | 10.163 (0.5%) |
| $1^3D_1$ | 9.72905 | 10.138 | 10.154 | 10.146 | -- | 10.145 | 10.149 | 10.147 | 10.117 | -- |
| $2^3D_3$ | 9.94704 | 10.455 | 10.449 | 10.436 | -- | 10.442 | 10.456 | 10.447 | 10.422 | -- |
| $2^1D_2$ | 9.94405 | 10.450 | 10.445 | 10.432 | -- | 10.439 | 10.452 | 10.440 | 10.419 | -- |
| $2^3D_2$ | 9.94259 | 10.449 | 10.443 | 10.432 | -- | 10.438 | 10.450 | 10.437 | 10.418 | -- |
| $2^3D_1$ | 9.93775 | 10.441 | 10.435 | 10.425 | -- | 10.432 | 10.443 | 10.428 | 10.414 | -- |
| $3^3D_3$ | 10.1435 | 10.711 | 10.717 | -- | -- | 10.680 | -- | 10.652 | -- | -- |
| $3^1D_2$ | 10.1405 | 10.706 | 10.713 | -- | -- | 10.677 | -- | 10.646 | --- | -- |

| Level | | | | | | | | | |
|---|---|---|---|---|---|---|---|---|---|
| $3^3D_2$ | 10.1391 | 10.705 | 10.711 | -- | -- | 10.676 | -- | 10.645 | -- | -- |
| $3^3D_1$ | 10.1344 | 10.698 | 10.704 | -- | -- | 10.670 | -- | 10.637 | -- | -- |
| $4^3D_3$ | 10.3284 | 10.939 | 10.963 | -- | -- | 10.886 | -- | 10.817 | -- | -- |
| $4^1D_2$ | 10.3255 | 10.935 | 10.959 | -- | -- | 10.883 | -- | 10.813 | -- | -- |
| $4^3D_2$ | 10.3241 | 10.934 | 10.957 | -- | -- | 10.882 | -- | 10.811 | -- | -- |
| $4^3D_1$ | 10.3195 | 10.928 | 10.949 | -- | -- | 10.877 | -- | 10.811 | -- | -- |
| $1^3F_2$ | 9.72948 | 10.350 | 10.343 | 10.338 | -- | -- | 10.353 | -- | 10.315 | -- |
| $1^3F_3$ | 9.7361 | 10.355 | 10.346 | 10.340 | 10.302 | -- | 10.356 | -- | 10.321 | -- |
| $1^1F_3$ | 9.73759 | 10.355 | 10.347 | 10.339 | 10.319 | -- | 10.356 | -- | 10.322 | -- |
| $1^3F_4$ | 9.74242 | 10.358 | 10.349 | 10.340 | -- | -- | 10.357 | -- | -- | -- |
| $2^3F_2$ | 9.93815 | 10.615 | 10.610 | -- | -- | -- | 10.610 | -- | -- | -- |
| $2^3F_3$ | 9.94462 | 10.619 | 10.614 | -- | -- | -- | 10.613 | -- | -- | -- |
| $2^1F_3$ | 9.94608 | 10.619 | 10.647 | -- | -- | -- | 10.613 | -- | -- | -- |
| $2^3F_4$ | 9.9508 | 10.622 | 10.617 | -- | -- | -- | 10.615 | -- | -- | -- |
| $3^3F_2$ | 10.1348 | 10.850 | -- | -- | -- | -- | -- | -- | -- | -- |
| $3^3F_3$ | 10.1411 | 10.853 | -- | -- | -- | -- | -- | -- | -- | -- |
| $3^1F_3$ | 10.1425 | 10.853 | -- | -- | -- | -- | -- | -- | -- | -- |
| $3^3F_4$ | 10.1471 | 10.856 | -- | -- | -- | -- | -- | -- | -- | -- |

Table 6. $B_c$ Meson mass spectrum of S and P-states in GeV

| Level | Present work | [27] | [61] | [69] | [76] | [16] | PDG [68] |
|---|---|---|---|---|---|---|---|
| $1^1S_0$ | 6.31528 | 6.272 | 6.278 | 6.272 | 6.271 | 6.275 | 6.275 (0.64%) |
| $1^3S_1$ | 6.62256 | 6.321 | 6.331 | 6.333 | 6.338 | 6.314 | -- |
| $2^1S_0$ | 6.85048 | 6.864 | 6.863 | 6.842 | 6.855 | 6.838 | 6.842 (0.124%) |
| $2^3S_1$ | 6.91694 | 6.900 | 6.873 | 6.882 | 6.887 | 6.850 | -- |
| $3^1S_0$ | 7.18692 | 7.306 | 7.244 | 7.226 | 7.250 | -- | -- |
| $3^3S_1$ | 7.18587 | 7.338 | 7.249 | 7.258 | 7.272 | -- | -- |
| $4^1S_0$ | 7.43221 | 7.684 | 7.564 | 7.585 | -- | -- | -- |

| State | | | | | | | |
|---|---|---|---|---|---|---|---|
| $4^3S_1$ | 7.43125 | 7.714 | 7.568 | 7.609 | -- | -- | -- |
| $5^1S_0$ | 7.65579 | 8.025 | 7.852 | 7.928 | -- | -- | -- |
| $5^3S_1$ | 7.6549 | 8.054 | 7.855 | 7.947 | -- | -- | -- |
| $6^1S_0$ | 7.85941 | 8.340 | 8.120 | -- | -- | -- | -- |
| $6^3S_1$ | 7.85861 | 8.368 | 8.122 | -- | -- | -- | -- |
| $1^3P_0$ | 6.60714 | 6.686 | 6.748 | 6.699 | 6.706 | 6.672 | -- |
| $1^3P_1$ | 6.62021 | 6.705 | 6.767 | 6.750 | 6.741 | 6.766 | -- |
| $1^1P_1$ | 6.62531 | 6.706 | 6.769 | 6.743 | 6.750 | 6.828 | -- |
| $1^3P_2$ | 6.6298 | 6.712 | 6.775 | 6.761 | 6.768 | 6.776 | -- |
| $2^3P_0$ | 6.90206 | 7.146 | 7.139 | 7.094 | 7.122 | 6.914 | -- |
| $2^3P_1$ | 6.91462 | 7.165 | 7.155 | 7.134 | 7.145 | 7.259 | -- |
| $2^1P_1$ | 6.91952 | 7.168 | 7.156 | 7.094 | 7.150 | 7.322 | -- |
| $2^3P_2$ | 6.92391 | 7.173 | 7.162 | 7.157 | 7.164 | 7.232 | -- |
| $3^3P_0$ | 7.17152 | 7.536 | 7.463 | 7.474 | -- | -- | -- |
| $3^3P_1$ | 7.18358 | 7.555 | 7.479 | 7.510 | -- | -- | -- |
| $3^1P_1$ | 7.18829 | 7.559 | 7.479 | 7.500 | -- | -- | -- |
| $3^3P_2$ | 7.19258 | 7.565 | 7.485 | 7.524 | -- | -- | -- |
| $4^3P_0$ | 7.41742 | 7.885 | -- | 7.817 | -- | -- | -- |
| $4^3P_1$ | 7.42899 | 7.905 | -- | 7.853 | -- | -- | -- |
| $4^1P_1$ | 7.43351 | 7.908 | -- | 7.844 | -- | -- | -- |
| $4^3P_2$ | 7.4377 | 7.915 | -- | 7.867 | -- | -- | -- |
| $5^3P_0$ | 7.64159 | 8.207 | -- | -- | -- | -- | -- |
| $5^3P_1$ | 7.65268 | 8.226 | -- | -- | -- | -- | -- |
| $5^1P_1$ | 7.65702 | 8.230 | -- | -- | -- | -- | -- |

Table 7. $B_c$ Meson mass spectrum of D and F-states in GeV

| Level | Present work | [27] | [61] | [69] | [76] | [16] |
|---|---|---|---|---|---|---|
| $1^3D_3$ | 6.63884 | 6.990 | 7.026 | 7.029 | 7.045 | 6.980 |
| $1^1D_2$ | 6.62836 | 6.994 | 7.035 | 7.026 | 7.041 | 7.009 |
| $1^3D_2$ | 6.62327 | 6.997 | 7.025 | 7.025 | 7.036 | 7.154 |
| $1^3D_1$ | 6.60632 | 6.998 | 7.030 | 7.021 | 7.028 | 7.078 |
| $2^3D_3$ | 6.93259 | 7.399 | 7.363 | 7.405 | -- | -- |
| $2^1D_2$ | 6.92242 | 7.401 | 7.370 | 7.400 | -- | -- |
| $2^3D_2$ | 6.91752 | 7.403 | 7.361 | 7,399 | -- | -- |
| $2^3D_1$ | 6.9012 | 7.403 | 7.365 | 7.392 | -- | -- |
| $3^3D_3$ | 7.20091 | 7.761 | -- | 7.750 | -- | -- |
| $3^1D_2$ | 7.19104 | 7.762 | -- | 7.743 | -- | -- |
| $3^3D_2$ | 7.18633 | 7.764 | -- | 7.741 | -- | -- |
| $3^3D_1$ | 7.17061 | 7.762 | -- | 7.732 | -- | -- |
| $4^3D_3$ | 7.44569 | 8.092 | -- | -- | -- | -- |
| $4^1D_2$ | 7.43611 | 8.093 | -- | -- | -- | -- |
| $4^3D_2$ | 7.43159 | 8.094 | -- | -- | -- | -- |
| $4^3D_1$ | 7.41647 | 8.091 | -- | -- | -- | -- |
| $1^3F_2$ | 6.60519 | 7.234 | -- | 7.273 | 7.269 | -- |
| $1^3F_3$ | 6.62785 | 7.242 | -- | 7.269 | 7.276 | -- |
| $1^1F_3$ | 6.63295 | 7.241 | -- | 7.268 | 7.266 | -- |
| $1^3F_4$ | 6.64949 | 7.244 | -- | 7.277 | 7.271 | -- |
| $2^3F_2$ | 6.90002 | 7.607 | -- | 7.618 | -- | -- |
| $2^3F_3$ | 6.92188 | 7.615 | -- | 7.616 | -- | -- |
| $2^1F_3$ | 6.92678 | 7.614 | -- | 7.615 | -- | -- |
| $2^3F_4$ | 6.94281 | 7.617 | -- | 7.617 | -- | -- |
| $3^3F_2$ | 7.1694 | 7.946 | -- | -- | -- | -- |
| $3^3F_3$ | 7.19049 | 7.954 | -- | -- | -- | -- |
| $3^1F_3$ | 7.19517 | 7.953 | -- | -- | -- | -- |
| $3^3F_4$ | 7.2107 | 7.956 | -- | -- | -- | -- |

## Conclusions

In the present work the quarkonium systems are described by solving nonrelativistic Schrödinger equation with relativistic corrections to obtain a quantitative description of these systems. There is an improvement of fitting with the available experimental data in addition to other theoretical states with the present work. In the future it is expected to compare with new experimental data .

## References

[1] C. S. Fischer, S. Kubrak, and R. Williams, "Spectra of heavy mesons in the Bethe-Salpeter approach," Eur. Phys. J. A, vol. 51, no. 1, pp. 1–9, 2015, doi: 10.1140/epja/i2015-15010-7.

[2] K. Nochi, T. Kawanai, and S. Sasaki, "Bethe-Salpeter wave functions of ηc (2S) and ψ (2S) states from full lattice QCD," Phys. Rev. D, vol. 94, no. 11, pp. 1–12, 2016, doi: 10.1103/PhysRevD.94.114514.

[3] V. Sauli, "Bethe-Salpeter study of radially excited vector quarkonia," Phys. Rev. D - Part. Fields, Gravit. Cosmol., vol. 86, no. 9, p. 096004, Nov. 2012, doi: 10.1103/PhysRevD.86.096004.

[4] A. N. Mitra, "Bethe-Salpeter equations for q q̄ and qqq systems in the instantaneous approximation," Zeitschrift für Phys. C Part. Fields, vol. 8, no. 1, pp. 25–31, 1981, doi: 10.1007/BF01429827.

[5] T. Hilger, C. Popovici, M. Gómez-Rocha, and A. Krassnigg, "Spectra of heavy quarkonia in a Bethe-Salpeter-equation approach," Phys. Rev. D - Part. Fields, Gravit. Cosmol., vol. 91, no. 3, pp. 30–34, 2015, doi: 10.1103/PhysRevD.91.034013.

[6] K. M. Maung, D. E. Kahana, and J. W. Norbury, "Solution of two-body relativistic bound-state equations with confining plus coulomb interactions," Phys. Rev. D, vol. 47, no. 3, pp. 1182–1189, Feb. 1993, doi: 10.1103/PhysRevD.47.1182.

[7] S. Leitão, A. Stadler, M. T. Peña, and E. P. Biernat, "Linear confinement in momentum space: Singularity-free bound-state equations," Phys. Rev. D - Part. Fields, Gravit. Cosmol., vol. 90, no. 9, p. 096003, Nov. 2014, doi:

10.1103/PhysRevD.90.096003.

[8]    C. McNeile, C. T. H. Davies, E. Follana, K. Hornbostel, and G. P. Lepage, "Heavy meson masses and decay constants from relativistic heavy quarks in full lattice QCD," Phys. Rev. D - Part. Fields, Gravit. Cosmol., vol. 86, no. 7, p. 074503, Oct. 2012, doi: 10.1103/PhysRevD.86.074503.

[9]    J. J. Dudek, R. G. Edwards, N. Mathur, and D. G. Richards, "Charmonium excited state spectrum in lattice QCD," Phys. Rev. D - Part. Fields, Gravit. Cosmol., vol. 77, no. 3, 2008, doi: 10.1103/PhysRevD.77.034501.

[10]   O. Lakhina and E. S. Swanson, "Dynamic properties of charmonium," Phys. Rev. D - Part. Fields, Gravit. Cosmol., vol. 74, no. 1, 2006, doi: 10.1103/PhysRevD.74.014012.

[11]   S. Meinel, "Bottomonium spectrum from lattice QCD with 2+1 flavors of domain wall fermions," Phys. Rev. D - Part. Fields, Gravit. Cosmol., vol. 79, no. 9, pp. 1–11, 2009, doi: 10.1103/PhysRevD.79.094501.

[12]   R. Nag, S. N. Mukherjee, and S. Sanyal, "Criterion for a unique non-relativistic quark model," Pramana J. Phys., vol. 32, no. 6, pp. 761–768, 1989, doi: 10.1007/BF02845997.

[13]   V. H. Zaveri, "Quarkonium and hydrogen spectra with spin-dependent relativistic wave equation," Pramana - J. Phys., vol. 75, no. 4, pp. 579–598, 2010, doi: 10.1007/s12043-010-0140-6.

[14]   J. F. Gunion and L. F. Li, "Relativistic treatment of the quark-confinement potential," Phys. Rev. D, vol. 12, no. 11, pp. 3583–3588, Dec. 1975, doi: 10.1103/PhysRevD.12.3583.

[15]   J. T. Laverty, S. F. Radford, and W. W. Repko, "γγ and gg Decay Rates for Equal Mass Heavy Quarkonia," no. 1, pp. 1–6, 2009, http://arxiv.org/abs/0901.3917.

[16]   A. P. Monteiro, M. Bhat, and K. B. V. Kumar, "C$\bar{b}$ Spectrum and Decay Properties With Coupled Channel Effects," Phys. Rev. D, vol. 95, no. 5, pp. 1–12, 2017, doi: 10.1103/PhysRevD.95.054016.

[17]   S. F. Radford and W. W. Repko, "Potential model calculations and predictions for heavy quarkonium," Phys. Rev. D - Part. Fields, Gravit. Cosmol., vol. 75, no. 7, pp. 1–9, 2007, doi: 10.1103/PhysRevD.75.074031.

[18]   J. Morishita, M. Kawaguchi, and T. Morii, "Spectroscopy of atomlike mesons Qq in a semirelativistic theory," Phys. Rev. D, vol. 37, no. 1, pp. 159–178, Jan. 1988, doi: 10.1103/PhysRevD.37.159.

[19]   S. N. Gupta, S. F. Radford, and W. W. Repko, "Semirelativistic Potential Model for Heavy Quarkonia," Phys.\ Rev.\ D, vol. 34, pp. 201–206, 1986, doi: 10.1103/PhysRevD.34.201.

[20]   A. K. Rai, B. Patel, and P. C. Vinodkumar, "Properties of Q$\bar{Q}$ mesons in nonrelativistic QCD formalism," Phys. Rev. C - Nucl. Phys., vol. 78, no. 5, p. 055202, Nov. 2008, doi: 10.1103/PhysRevC.78.055202.

[21]   B. Patel and P. C. Vinodkumar, "Properties of QQ(Q ∈ b, c) mesons in Coulomb plus power potential (CPPv)," J. Phys. G Nucl. Part. Phys., vol. 36, no. 3, 2009, doi: 10.1088/0954-3899/36/3/035003.

[22]   H. Mansour, A. Gamal, and M. Abolmahassen, "Spin Splitting Spectroscopy of Heavy

Quark and Antiquarks Systems," Adv. High Energy Phys., vol. 2020, p. 2356436, 2020, doi: 10.1155/2020/2356436.

[23] R. Nag, S. Sanyal, and S. N. Mukherjee, "Electromagnetic structure of the proton and baryon spectrum in the nonrelativistic quark model," Phys. Rev. D, vol. 36, no. 9, pp. 2788–2799, Nov. 1987, doi: 10.1103/PhysRevD.36.2788.

[24] A. Maireche, "New nonrelativistic quarkonium masses in the two-dimensional space-phase using Bopp's shift method and standard perturbation theory," J. Nano- Electron. Phys., vol. 9, no. 6, 2017, doi: 10.21272/jnep.9(6).06006.

[25] A. D. Antia, "Solutions of Nonrelativistic Schrödinger Equation with Scarf II Plus Rosen-Morse II Potential via Ansaltz Method," Am. J. Phys. Chem., vol. 4, no. 5, p. 38, 2015, doi: 10.11648/j.ajpc.20150405.11.

[26] E. Yazdankish, "Solving of the Schrodinger equation analytically with an approximated scheme of the Woods-Saxon potential by the systematical method of Nikiforov-Uvarov," Int. J. Mod. Phys. E, vol. 29, no. 6, 2020, doi: 10.1142/S0218301320500329.

[27] N. R. Soni, B. R. Joshi, R. P. Shah, H. R. Chauhan, and J. N. Pandya, "$QQ^-$ ($Q\in$ {b, c}) spectroscopy using the Cornell potential," Eur. Phys. J. C, vol. 78, no. 7, 2018, doi: 10.1140/epjc/s10052-018-6068-6.

[28] H. Mansour and A. Gamal, "Bound State of Heavy Quarks Using a General Polynomial Potential," Adv. High Energy Phys., vol. 2018, no. 2, 2018, doi: 10.1155/2018/7269657.

[29] B. I. Ita et al., "Bound State Solutions of the Schrodinger Equation for the More General Exponential Screened Coulomb Potential Plus Yukawa (MGESCY) Potential Using Nikiforov-Uvarov Method," J. Quantum Inf. Sci., vol. 08, no. 01, pp. 24–45, 2018, doi: 10.4236/jqis.2018.81003.

[30] G. C. Joshi and J. W. G. Wignall, "Universal potential curves for quarkonia," Lett. al Nuovo Cim., vol. 35, no. 14, pp. 437–442, 1982, doi: 10.1007/BF02906872.

[31] S. M. Kuchin and N. V Maksimenko, "Theoretical Estimations of the Spin – Averaged Mass Spectra of Heavy Quarkonia and Bc Mesons," vol. 1, no. 3, pp. 295–298, 2013, doi: 10.13189/ujpa.2013.010310.

[32] B. I. Ita and A. I. Ikeuba, "Solutions of the Dirac Equation with Gravitational plus Exponential Potential," Appl. Math., vol. 04, no. 10, pp. 1–6, 2013, doi: 10.4236/am.2013.410a3001.

[33] B. I. Ita, C. O. Ehi-Eromosele, A. Edobor-Osoh, A. I. Ikeuba, and T. A. Anake, "Solutions of the Schrödinger equation with inversely quadratic effective plus Mie-type potential using Nikiforov-Uvarov method," in AIP Conference Proceedings, 2014, vol. 1629, no. November, pp. 235–238, doi: 10.1063/1.4902278.

[34] A. P. Monteiro and M. Bhat, "Analytical solutions of the Schrodinger equation with the Woods-Saxon potential for l = 0 states," vol. 60, pp. 664–665, 2015.

[35] A. Vega and F. Rojas, "Confinement potentials for the study of heavy mesons," 2018, http://arxiv.org/abs/1810.08080.

[36] A. K. Rai and P. C. Vinodkumar, "Properties of Bc meson," Pramana - J. Phys., vol. 66, no. 5, pp. 953–958, 2006, doi: 10.1007/BF02704795.

[37]   H. Mansour and A. Gamal, "Two body problems with magnetic interactions," vol. 9, no. 2, pp. 51–58, 2019, doi: 10.5923/j.jnpp.20190902.02.

[38]   H. Mansour and A. Gamal, "Spectroscopy of the Quarkonium Systems for Heavy Quarks," vol. 10, no. 1, pp. 13–22, 2020, doi: 10.5923/j.jnpp.20201001.03.

[39]   J. N. Pandya, N. R. Soni, N. Devlani, and A. K. Rai, "Decay rates and electromagnetic transitions of heavy quarkonia," Chinese Phys. C, vol. 39, no. 12, pp. 1–12, 2015, doi: 10.1088/1674-1137/39/12/123101.

[40]   S. Laachir and A. Laaribi, "Exact Solutions of the Helmholtz equation via the Nikiforov-Uvarov Method," Nov. 2013, doi: 10.5281/ZENODO.1089211.

[41]   B. Ita, P. Tchoua, E. Siryabe, and G. E. Ntamack, "Solutions of the Klein-Gordon Equation with the Hulthen Potential Using the Frobenius Method," vol. 4, no. 5, pp. 173–177, 2014, doi: 10.5923/j.ijtmp.20140405.02.

[42]   F. Yaşuk, C. Berkdemir, and A. Berkdemir, "Exact solutions of the Schrödinger equation with non-central potential by the Nikiforov-Uvarov method," J. Phys. A. Math. Gen., vol. 38, no. 29, pp. 6579–6586, 2005, doi: 10.1088/0305-4470/38/29/012.

[43]   B. J. Falaye, K. J. Oyewumi, and M. Abbas, "Exact solution of Schrödinger equation with q-deformed quantum potentials using Nikiforov - Uvarov method," Chinese Phys. B, vol. 22, no. 11, 2013, doi: 10.1088/1674-1056/22/11/110301.

[44]   S. M. Ikhdair, " Approximate l -States of the Manning-Rosen Potential by Using Nikiforov-Uvarov Method ," ISRN Math. Phys., vol. 2012, pp. 1–20, 2012, doi: 10.5402/2012/201525.

[45]   H. Karayer, D. Demirhan, and F. Büyükkiliç, "Extension of nikiforov-uvarov method for the solution of Heun equation," J. Math. Phys., vol. 56, no. 6, 2015, doi: 10.1063/1.4922601.

[46]   C. Berkdemir, A. Berkdemir, and R. Sever, "Polynomial solutions of the Schrödinger equation for the generalized Woods-Saxon potential," Phys. Rev. C - Nucl. Phys., vol. 72, no. 2, p. 027001, Aug. 2005, doi: 10.1103/PhysRevC.72.027001.

[47]   H. Goudarzi and V. Vahidi, "Supersymmetric approach for eckart potential using the NU method," Adv. Stud. Theor. Phys., vol. 5, no. 9–12, pp. 469–476, 2011.

[48]   H. Karayer, D. Demirhan, and F. Büyükkiliç, "Conformable Fractional Nikiforov - Uvarov Method," Commun. Theor. Phys., vol. 66, no. 1, pp. 12–18, 2016, doi: 10.1088/0253-6102/66/1/012.

[49]   A. F. Al-Jamel and H. Widyan, "Heavy Quarkonium Mass Spectra in A Coulomb Field Plus Quadratic Potential Using Nikiforov-Uvarov Method," Appl. Phys. Res., vol. 4, no. 3, pp. 94–99, 2012, doi: 10.5539/apr.v4n3p94.

[50]   M. Abu-Shady, "Quarkonium masses in a hot QCD medium using conformable fractional of the Nikiforov-Uvarov method," Int. J. Mod. Phys. A, pp. 1–15, 2019, doi: 10.1142/S0217751X19502014.

[51]   T. Burch et al., "Quarkonium mass splittings in three-flavor lattice QCD," Phys. Rev. D - Part. Fields, Gravit. Cosmol., vol. 81, no. 3, pp. 1–21, 2010, doi: 10.1103/PhysRevD.81.034508.

[52]   T. Liu, A. A. Penin, and A. Rayyan, "Coulomb artifacts and bottomonium hyperfine splitting in lattice NRQCD," J. High Energy Phys., vol. 2017, no. 2, 2017, doi:

10.1007/JHEP02(2017)084.

[53] J. F. Gunion and L. F. Li, "Relativistic treatment of the quark-confinement potential," Phys. Rev. D, vol. 12, no. 11, pp. 3583–3588, Dec. 1975, doi: 10.1103/PhysRevD.12.3583.

[54] E. Eichten and F. Feinberg, "Spin-dependent forces in quantum chromodynamics," Phys. Rev. D, vol. 23, no. 11, pp. 2724–2744, 1981, doi: 10.1103/PhysRevD.23.2724.

[55] I. Haysak, Y. Fekete, V. Morokhovych, S. Chalupka, and M. Salak, "Spin effects in two quark system and mixed states," Czechoslov. J. Phys., vol. 55, no. 5, pp. 541–554, 2005, doi: 10.1007/s10582-005-0059-1.

[56] N. Isgur and G. Karl, "P-wave baryons in the quark model," Phys. Rev. D, vol. 18, no. 11, pp. 4187–4205, 1978, doi: 10.1103/PhysRevD.18.4187.

[57] S. S. Gershtein, V. V. Kiselev, A. K. Likhoded, and A. V. Tkabladze, "Bc spectroscopy," Phys. Rev. D, vol. 51, no. 7, pp. 3613–3627, 1995, doi: 10.1103/PhysRevD.51.3613.

[58] N. Devlani and A. K. Rai, "Spectroscopy and decay properties of B and Bs mesons," Eur. Phys. J. A, vol. 48, no. 7, p. 104, 2012, doi: 10.1140/epja/i2012-12104-8.

[59] S. Patel, P. C. Vinodkumar, and S. Bhatnagar, "Decay rates of charmonia within a quark-antiquark confining potential," Chinese Phys. C, vol. 40, no. 5, 2016, doi: 10.1088/1674-1137/40/5/053102.

[60] W. Kwong and J. L. Rosner, "D-wave quarkonium levels of the family," Phys. Rev. D, vol. 38, no. 1, pp. 279–297, 1988, doi: 10.1103/PhysRevD.38.279.

[61] N. Devlani, V. Kher, and A. K. Rai, "Masses and electromagnetic transitions of the Bc mesons," Eur. Phys. J. A, vol. 50, no. 10, pp. 627–631, 2014, doi: 10.1140/epja/i2014-14154-2.

[62] D. B. Lichtenberg and J. G. Wills, "Spectrum of Strange-Quark-Antiquark Bound States," Phys. Rev. Lett., vol. 35, no. 16, pp. 1055–1059, Oct. 1975, doi: 10.1103/PhysRevLett.35.1055.

[63] S. F. Radford and W. W. Repko, "Hyperfine splittings in the $b\bar{b}$ system," Nucl. Phys. A, vol. 865, no. 1, pp. 69–75, 2011, doi: 10.1016/j.nuclphysa.2011.06.032.

[64] H. C. Bolton and A. R. Edmonds, "Angular Momentum in Quantum Mechanics," Math. Gaz., vol. 43, no. 344, p. 157, 1959, doi: 10.2307/3610250.

[65] T. J. Burns, "Hyperfine splitting and the experimental candidates for ηb(2S)," Phys. Rev. D - Part. Fields, Gravit. Cosmol., vol. 87, no. 3, pp. 1–8, 2013, doi: 10.1103/PhysRevD.87.034022.

[66] Y. Kiyo and Y. Sumino, "Perturbative heavy quarkonium spectrum at next-to-next-to-next-to-leading order," Phys. Lett. Sect. B Nucl. Elem. Part. High-Energy Phys., vol. 730, pp. 76–80, 2014, doi: 10.1016/j.physletb.2014.01.030.

[67] E. Eichten, S. Godfrey, H. Mahlke, and J. L. Rosner, "Quarkonia and their transitions," Rev. Mod. Phys., vol. 80, no. 3, pp. 1–80, 2008, doi: 10.1103/RevModPhys.80.1161.

[68] C. Patrignani et al., "Review of particle physics," Chinese Physics C, vol. 40, no. 10. 2016, doi: 10.1088/1674-1137/40/10/100001.

[69] D. Ebert, R. N. Faustov, and V. O. Galkin, "Spectroscopy and Regge trajectories of heavy quarkonia and Bc mesons," Eur. Phys. J. C, vol. 71, no. 12, pp. 1–13, 2011, doi: 10.1140/epjc/s10052-011-1825-9.

[70] W. J. Deng, H. Liu, L. C. Gui, and X. H. Zhong, "Charmonium spectrum and electromagnetic transitions with higher multipole contributions," Phys. Rev. D, vol. 95, no. 3, pp. 1–20, 2017, doi: 10.1103/PhysRevD.95.034026.

[71] B. Q. Li and K. T. Chao, "Higher charmonia and X, Y, Z states with screened potential," Phys. Rev. D - Part. Fields, Gravit. Cosmol., vol. 79, no. 9, 2009, doi: 10.1103/PhysRevD.79.094004.

[72] M. Shah, A. Parmar, and P. C. Vinodkumar, "Leptonic and digamma decay properties of S-wave quarkonia states," Phys. Rev. D - Part. Fields, Gravit. Cosmol., vol. 86, no. 3, p. 034015, Aug. 2012, doi: 10.1103/PhysRevD.86.034015.

[73] T. Barnes, S. Godfrey, and E. S. Swanson, "Higher charmonia," Phys. Rev. D - Part. Fields, Gravit. Cosmol., vol. 72, no. 5, pp. 1–28, 2005, doi: 10.1103/PhysRevD.72.054026.

[74] S. Godfrey and K. Moats, "Bottomonium mesons and strategies for their observation," Phys. Rev. D - Part. Fields, Gravit. Cosmol., vol. 92, no. 5, p. 054034, Sep. 2015, doi: 10.1103/PhysRevD.92.054034.

[75] J. Segovia, P. G. Ortega, D. R. Entem, and F. Fernández, "Bottomonium spectrum revisited," Phys. Rev. D, vol. 93, no. 7, 2016, doi: 10.1103/PhysRevD.93.074027.

[76] S. Godfrey, "Spectroscopy of Bc mesons in the relativized quark model," Phys. Rev. D - Part. Fields, Gravit. Cosmol., vol. 70, no. 5, p. 15, Sep. 2004, doi: 10.1103/PhysRevD.70.054017.

[77] B. Q. Li and K. T. Chao, "Bottomonium spectrum with screened potential," Commun. Theor. Phys., vol. 52, no. 4, pp. 653–661, 2009, doi: 10.1088/0253-6102/52/4/20.